\documentclass[preprintnumbers,twocolumn,groupedaddress,nofootinbib,aps]{revtex4}

\usepackage{amsmath,amssymb}
\usepackage{graphicx}
\usepackage{color}
\usepackage{hyperref}
\usepackage{url}
\usepackage{slashed}
\usepackage{subfigure}
\usepackage{slashed,bbm}
\usepackage{graphics,psfrag,epsfig}
\usepackage{dsfont}

\newcommand{\spitz}[1]{ \left\langle  #1 \right\rangle }

\newcommand{\pr}[1]{{{#1}^{\prime}}}

\begin{document}


\title{Mott multicriticality of Dirac electrons in graphene}

\author{Laura Classen$^1$,
	    Igor F. Herbut$^2$,
            Lukas Janssen$^2$,
            Michael M. Scherer$^1$}

\affiliation{$^1$Institut f\"ur Theoretische Physik, Universit\"at Heidelberg, D-69120, Germany}
\affiliation{$^2$Department of Physics, Simon Fraser University, Burnaby, British Columbia, Canada V5A 1S6}

\begin{abstract}
We study the multicritical behavior for the semimetal-insulator transitions on graphene's honeycomb lattice using the Gross-Neveu-Yukawa effective theory with two order parameters: the SO(3) (Heisenberg) order parameter describes the antiferromagnetic transition, and the $\mathbb{Z}_2$ (Ising) order parameter describes the transition to a staggered density state.
Their coupling induces multicritical behavior which determines the structure of the phase diagram close to the multicritical point.
Depending on the number of fermion flavors $N_f$ and working in the perturbative regime in vicinity of three (spatial) dimensions, we observe first order or continuous phase transitions at the multicritical point.
For the graphene case of $N_f=2$ and within our low order approximation, the phase diagram displays a tetracritical structure.
\end{abstract}

\maketitle


\section{Introduction}

Graphene~\cite{castroneto2009} features a number of unconventional electronic properties which to a large extent can be captured by means of a simple tight-binding description of the electrons on the honeycomb lattice~\cite{semenoff}.
The experimental findings such as, for example,  the half-integer quantum Hall effect~\cite{novoselov2005a} or the Klein paradox~\cite{katsnelson2006a} suggest that this single-particle picture provides a very good starting point for the theoretical description of the material.
Electron-electron interactions therefore are often expected to play only a quantitative and not a qualitative role.
In fact, on the honeycomb lattice at charge neutrality the density of states vanishes linearly for energies close to the Fermi level, and an interaction-induced transition toward an ordered -- possibly Mott insulating -- state appears only when a minimal critical value of the interaction strength is exceeded~\cite{sorella1992,khvesh2001,gorbar2002,herbut2006}.
Depending on the interaction profile, i.e. the precise ratios of onsite, nearest-neighbor, and further interaction parameters, a great variety of different spontaneously broken symmetries can occur~\cite{herbut2008,honerkamp2008,herbut2009,raghu2008,grushin2013,daghofer2013, duric,doniach2007}.
The symmetry breaking patterns, most prominently, include chiral symmetry breaking phases, such as the antiferromagnetic spin density wave (SDW) or a charge density wave (CDW)  \cite{herbut2009,araki2012,wu2013,janssen2014}. However, more exotic states of matter such as quantum Hall and the quantum spin Hall phase
\cite{raghu2008,daghofer2013,grushin2013}, or the existence of a quantum spin liquid  have been discussed \cite{meng,sorella2012}.

Recent \emph{ab initio} calculations\cite{wehling2011,ulybyshev2013}, on the other hand, find values for the interaction parameters of graphene which seem to be close to a transition toward a spontaneously symmetry-broken phase.
This motivates a consideration of circumstances under which graphene could be driven through a quantum phase transition by tuning of external parameters.
One possibility could be a uniform mechanical stretching of the graphene sheet to increase the ratio of onsite interaction to nearest-neighbor hopping parameter.
The ratio between the onsite and the nearest-neighbor interaction might also be in principle tuned by proximity of a substrate that induces a screening of the Coulomb interaction.

\begin{figure}[b!]
\centering
 \includegraphics[width=0.31\textheight]{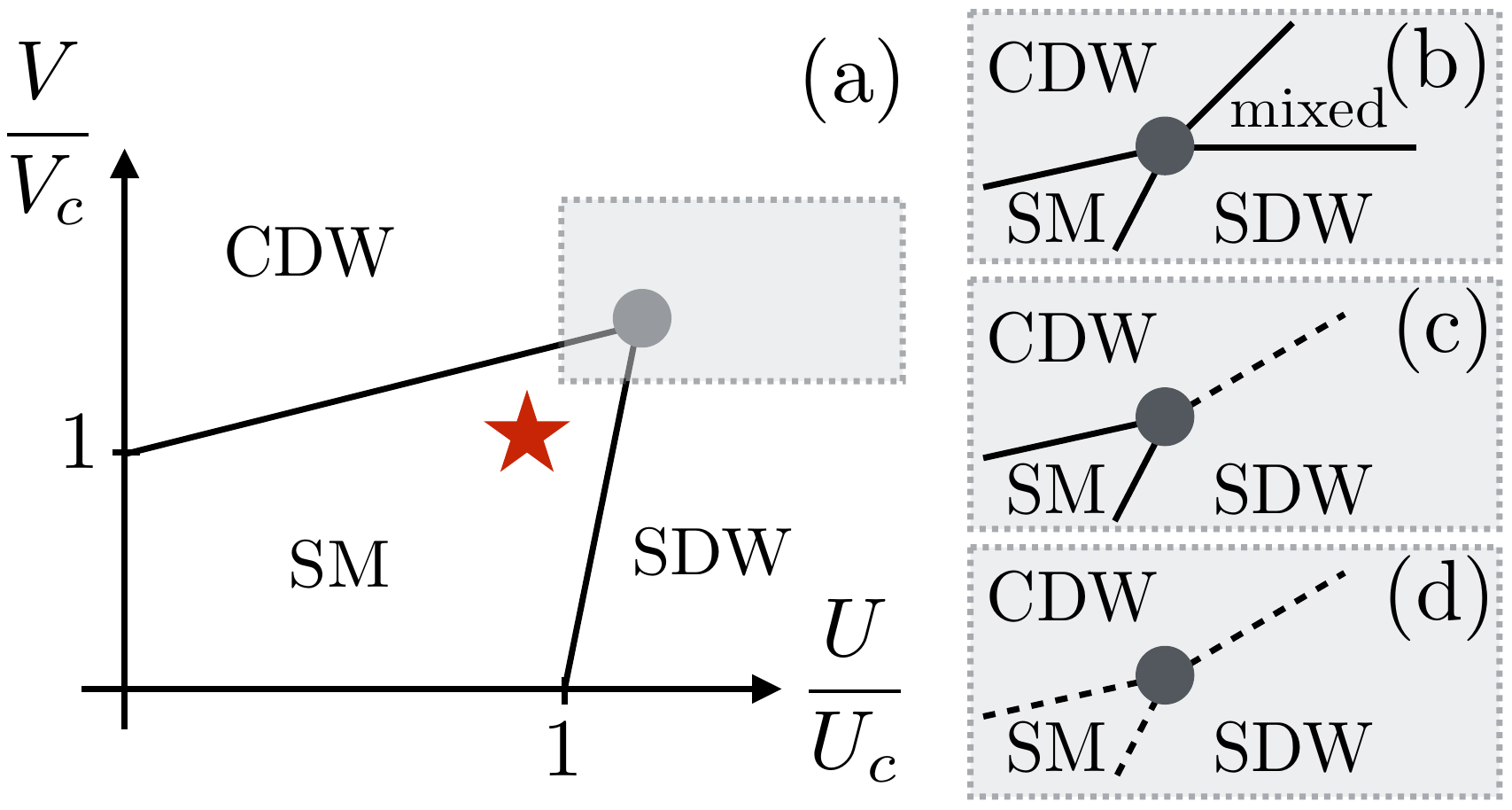}
 \caption{(a) Schematic phase diagram of the extended Hubbard model on the honeycomb lattice with onsite interaction $U$ and nearest-neighbor interaction $V$. Neutral suspended graphene is found to be in the semi-metallic state indicated by the star. The neighborhood of the multicritical point (grey shaded area) may be governed by a (b) tetra-critical or (c) bicritical structure or by a (d) first-order multicritical point. Solid lines denote second order and dashed lines first order phase transitions.}
\label{fig:phasediag}
\end{figure}

Here, we consider a situation with both a large onsite interaction, and a sizable nearest-neighbor interaction.
Separately, these interactions would trigger phase transitions toward the ordered SDW and the CDW states, respectively.
The experimental findings together with the \emph{ab initio} parameters suggest that graphene is close, but somewhat below the critical values for the formation of one of these ordered states, cf. Fig.~\ref{fig:phasediag}.
%

The separate critical behavior of these two orders for Dirac electrons on graphene's honeycomb lattice has been investigated in many studies; see Refs.~\onlinecite{herbut2009,janssen2014,assaad,wang}, for example.
On the other hand, with both interaction parameters being close to criticality, the system is close to the point where the transition lines into the SDW and the CDW phases meet, and multicritical behavior results.
The problem of competing order parameters is a complex many-body problem, and multicritical behavior is important to many different fields of physics. A prominent example arises in the field of correlated electrons, e.g. in the study of high-temperature superconductors and related complex materials. In graphene, multicritical behavior for the case of broken spin rotation symmetry has been studied \cite{roy2011}.
At the same time the study of low-dimensional Dirac electrons and their competing interactions can shed light on related structures in the phase diagram of quantum chromodynamics.

In this paper, we study the multicritical behavior of an effective theory for the semimetal-insulator transitions on graphene's honeycomb lattice, which assumes the form of a Gross-Neveu-Yukawa field theory with two coupled order parameters. We consider a generalized theory of this type with an arbitrary number of fermion flavors $N_f$.
The SO(3) order parameter describes the antiferromagnetic transition and the $\mathbb{Z}_2$ order parameter describes the transition to a staggered density state.
Their coupling induces multicritical behavior which determines the structure of the phase diagram close to the multicritical point \cite{nelson1974}.
A first order transition at the critical point is possible, or if the transition is continuous, potential bi- or tetracritical behavior decides over the existence of a mixed phase in the symmetry-broken regime (see inset in Fig.~\ref{fig:phasediag}).
We set up an $\epsilon$ expansion close to three (spatial) plus one (time) dimensions to study which one of these three possibilities is realized for electrons on the honeycomb lattice, and, in particular, investigate the dependence of the result on the number of fermion flavors $N_f$.
For small flavor number, to the first  order in $\epsilon$, we find that  the multicritical behavior is governed by the decoupled fixed point that consists of the chiral Heisenberg and the Ising fixed points from the two separate descriptions of the Mott insulator transitions.
This implies tetracritical behavior and a decoupled mixed phase of the two order parameters.
For intermediate fermion flavor numbers, all transitions are of first order. Finally, for a large number of flavors the phase diagram displays a bicritical structure with a first order transition between the CDW and SDW state.

The remainder of the paper is organized as follows. In the next section we present the effective model with two dynamical order parameters.
We then compute the fixed-point structure as a function of $N_f$ within first-order $\epsilon$ expansion in Sec.~\ref{sec:eps-expansion}.
In Sec.~\ref{sec:discussion} we discuss the resulting phase diagram and give concluding remarks.


\section{Extended Hubbard model on the honeycomb lattice}

To describe the behavior of electrons on the honeycomb lattice we start with the tight-binding Hamiltonian supplemented by the interaction terms,  $H=H_0+H_{\text{int}}$,
with
\begin{align}
	H_0&=-t\sum_{\vec{R},\vec{\delta}_i,s}u_s^\dagger(\vec{R}) v_s(\vec{R}+\vec{\delta}_i)+\text{h.c.}\,,\\
   H_{\text{int}}&=U\sum_{i}n_{i,\uparrow}n_{i,\downarrow}+V\sum_{\substack{\spitz{i,j},s,s'}}n_{i,s}n_{j,\pr{s}} \,,
\end{align}
where $u_s$ and $v_s$ are the electron annihilation operators
at the two triangular sublattices of the honeycomb lattice with spin projection $s=\ \uparrow, \downarrow$
and $\vec{R}$ denotes the sites of one triangular sublattice.
The $\vec{\delta}_i$ are the vectors to the three nearest neighbors on the second sublattice.
Explicitly, the position vectors of the bipartite lattice are spanned by
$
\vec{R}_1^T=a\left(\sqrt{3}/2,-1/2\right)
$
and
$
\vec{R}_2^T=a\left(0,1\right),
$
where $a$ is the lattice spacing which in the following is set to $a=1$. The second sublattice is generated by
$
\vec{R}+\vec{\delta}_i
$
with the three nearest-neighbor vectors
$
\vec{\delta}_1^T=\left(1/(2\sqrt{3}),1/2\right)$, $\vec{\delta}_2^T=\left(1/(2\sqrt{3}),-1/2\right)$, and $\vec{\delta}_3^T=\left(1/\sqrt{3},0\right)
$.
Then, for the non-interacting part of the Hamiltonian, the energy spectrum is described by two bands
$
\epsilon_{\vec{k}}=\pm t|\sum_{i=1}^{3}\exp({\vec{k}\cdot\vec{\delta}_i})|
$
which are linear and isotropic close to zero energy near the $K,K'$ points at the border of the Brillouin zone, where $\vec{K}^T=(2\pi/\sqrt{3},2\pi/3)$ and $\vec{K}'=-\vec{K}$.
Employing only Fourier-modes near the $K,K'$ points the continuum low-energy effective action corresponding to $H_0$ can be written as~\cite{semenoff}
\begin{align}
	S_F=\int_0^{1/T}d\tau d\vec{x}^{D-1}\left[\bar\Psi\left(\mathbbm{1}_2\otimes\gamma_\mu\right)\partial_\mu\Psi\right]
\end{align}
where $D$ is the space-time dimension and the spin-$\frac{1}{2}$ electrons are described by the 8-component Dirac field $\Psi=\left(\Psi_\uparrow,\Psi_\downarrow\right)^T$ with its conjugate $\bar\Psi=\Psi^\dagger(\mathbbm{1}_2\otimes\gamma_0)$ in $2<D<4$. Further, we have $\partial_\mu=(\partial_\tau, \vec{\nabla})$ and the Clifford algebra $\{\gamma_\mu,\gamma_\nu\}=2\delta_{\mu\nu}$ with $(4\times4)$-$\gamma$ matrices and we assume the summation convention over repeated indices, and $\mu,\nu = 0,\dots,D-1$. Explicitly, in (2+1)D we use a representation where
$\gamma_0=\mathbbm{1}_2\otimes\sigma_z$, $\quad \gamma_1=\sigma_z\otimes\sigma_y$, $\quad \gamma_{2}=\mathbbm{1}_2\otimes\sigma_x$.
With these definitions the relation between the Grassmann fields $u,v$ and $\Psi$ can be given as~\cite{herbut2006}
%
\begin{align}
	 \Psi_s^\dagger(\vec{x},\tau)&=T\sum_{\omega_n}\hspace{-0.12cm}\int^\Lambda\hspace{-0.2cm}\frac{d^{D-1}\vec{q}}{(2\pi)^{D-1}}e^{i\omega_n\tau+i\vec{q}\cdot\vec{x}}\big[
	u_s^\dagger(\vec{K}+\vec{q},\omega_n),\nonumber\\
	&\hspace{-1cm}v_s^\dagger(\vec{K}+\vec{q},\omega_n),u_s^\dagger(-\vec{K}+\vec{q},\omega_n),v_s^\dagger(-\vec{K}+\vec{q},\omega_n)
	\big].\label{eq:spinor}
\end{align}
%
The reference frame is chosen to be such that $q_x=\vec{q}\cdot\vec{K}/|\vec{K}|$ and $q_y=(\vec{K}\times\vec{q})\times\vec{K}/|\vec{K}|^2$ and further we set $\hbar=k_B=v_F=1$.
We also define the two additional $4\times4$ matrices that anticommute with all $\gamma_\mu$, namely $\gamma_3=\sigma_x\otimes\sigma_y$ and $\gamma_5=\sigma_y\otimes\sigma_y$.
Then $\gamma_{35}=-i\gamma_3\gamma_5$ commutes with all $\gamma_\mu$ and anticommutes with $\gamma_3$ and $\gamma_5$.
In the following, we will also consider the generalization to arbitrary number of Dirac points in the spectrum. Formally, this can be done by replacing $u_s(\pm \vec K + \vec q, \omega_n) \mapsto u_s(\pm\vec K_i + \vec q, \omega_n)$ and $v_s(\pm \vec K + \vec q, \omega_n) \mapsto v_s(\pm \vec K_i + \vec q, \omega_n)$ with $\vec K_i$ the position of the Dirac points, $i = 1,\dots,N_f$. We will refer to $N_f$ as the number of fermion flavors with $N_f=2$ being the physical case. For general $N_f$, the generalization of Eq.~\eqref{eq:spinor}, thus renders the fermion field $\Psi$ to have $2N_f$ spinorial components for each spin direction.
%


\paragraph*{Order Parameters and Interactions}

There are several different order parameters that induce various symmetry breaking patterns, e.g., time-reversal symmetry (TRS) breaking and chiral symmetry (CS) breaking.
In this work, we will consider CS breaking while TRS is preserved.
The CS breaking order parameter can be written as
\begin{align}
	\Phi=\big(\chi,\vec{\phi}\big)=\left(\langle \bar\Psi\Psi\rangle,\langle \bar\Psi(\vec{\sigma}\otimes\mathbbm{1}_{2N_f})\Psi\rangle\right)\,.
\end{align}
%
%
The Ising field $\chi$ is a spin singlet, and corresponds to a staggered density, or a charge density wave state,
$\chi \sim u_s^\dagger u_s - v_s^\dagger v_s$,
which can be triggered by a large nearest-neighbor density-density interaction.
The Heisenberg field $\vec{\phi}$ is a triplet and corresponds to a staggered magnetization or an antiferromagnetic spin density wave state,
$\vec\phi \sim u_s^\dagger \vec \sigma_{ss'} u_s + v_s^\dagger \vec \sigma_{ss'} v_s$, and it is triggered by a strong onsite interaction.
The order parameters which appear in the form of fermion bilinears can be promoted to be dynamical fields corresponding to a bosonic action
\begin{align}
	\hspace{-0.2cm}S_B&\hspace{-0.1cm}=\hspace{-0.1cm}\int\hspace{-0.1cm} d\tau d\vec{x}^{D-1}\Big[
	\frac{1}{2}\chi(-\partial_\mu^2+m_\chi^2)\chi+\frac{1}{2}\vec{\phi}\cdot(-\partial_\mu^2+m_\phi^2)\vec{\phi}\nonumber\\
	&\quad\quad\quad\quad\quad+\lambda_{\chi}\chi^4+\lambda_{\phi}\big(\vec{\phi}\cdot \vec{\phi}\big)^2+\lambda_{\chi\phi}\chi^2\vec{\phi}^2
	\Big]\,.
\end{align}
Finally, there are also Yukawa terms that couple bosonic and fermionic degrees of freedom
\begin{align}
	S_Y\hspace{-0.1cm}&=\hspace{-0.1cm}\int\hspace{-0.1cm} d\tau d\vec{x}^{D-1}\left[
	g_{\chi}\chi\bar\Psi (\mathds{1}_2\otimes\mathds{1}_{2N_f})\Psi\right.\\
	&\left.\hspace{2cm}+\ g_{\phi}\vec{\phi}\cdot\bar
	\Psi({\vec{\sigma}}\otimes\mathds{1}_{2N_f})\Psi
	\right].\nonumber
\end{align}
The complete action then is given by
$
S=S_F+S_B+S_Y.
$
Its form is dictated by the spin-rotational, time-reversal and sublattice-exchange symmetries.
This effective theory is explicitly relativistic, cf.\ Refs.~\cite{herbut2009,janssen2014}.
The Coulomb repulsion does not appear explicitly, while it may tune the transition entering through the nearest-neighbor repulsion.
Its long-range tail has been shown to be an irrelevant perturbation at the critical points~\cite{herbut2009,Juricic2009}.


\section{$\epsilon$ expansion}
\label{sec:eps-expansion}

For the above action $S$ and at zero temperature we can calculate the renormalization group equations in the Wilsonian scheme by simultaneously integrating out the fermionic as well as the bosonic modes within the narrow momentum shell $\Lambda/b<(\omega^2+\vec{q}^2)<\Lambda$.
At one-loop order in $D=4-\epsilon$ dimensions we find the following equations for the two Yukawa couplings
\begin{align}
	\frac{dg_\chi^2}{d\ln b}&=\epsilon g_\chi^2-(3+2 N_f)g_\chi^4-9g_\chi^2g_\phi^2\,,\\
	\frac{dg_\phi^2}{d\ln b}&=\epsilon g_\phi^2-(1+2 N_f)g_\phi^4-3g_\chi^2g_\phi^2\,,
\end{align}
where we have rescaled $g_i^2\rightarrow g_i^2/(8\pi^2\Lambda^\epsilon)$.
Additionally, in the $\epsilon$ expansion we obtain RG flow equations for the bosonic masses $m_\chi^2, m_\phi^2$ and the bosonic couplings $\lambda_\chi, \lambda_\phi, \lambda_{\chi\phi}$ that will be displayed below.
The fixed points of this set of equations and their surrounding will determine the properties of the multicritical point.

\begin{figure*}[t]
\centering
 \includegraphics{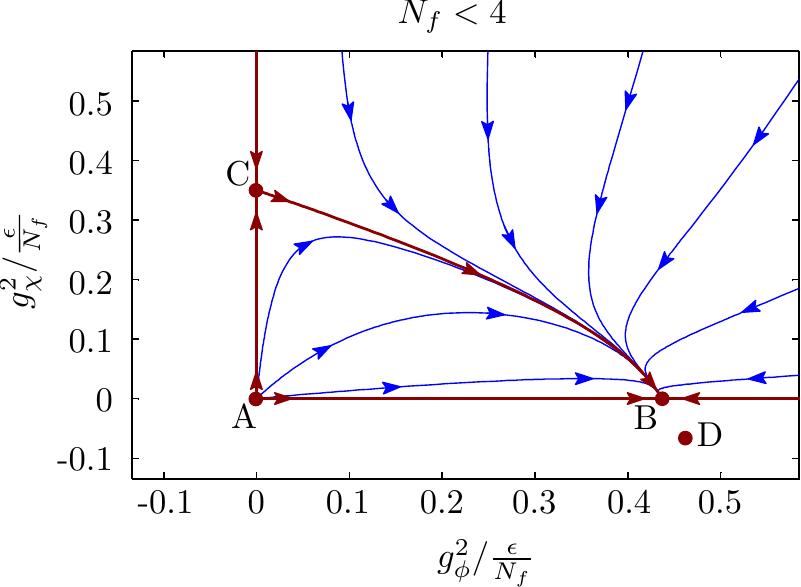} \hfill
 \includegraphics{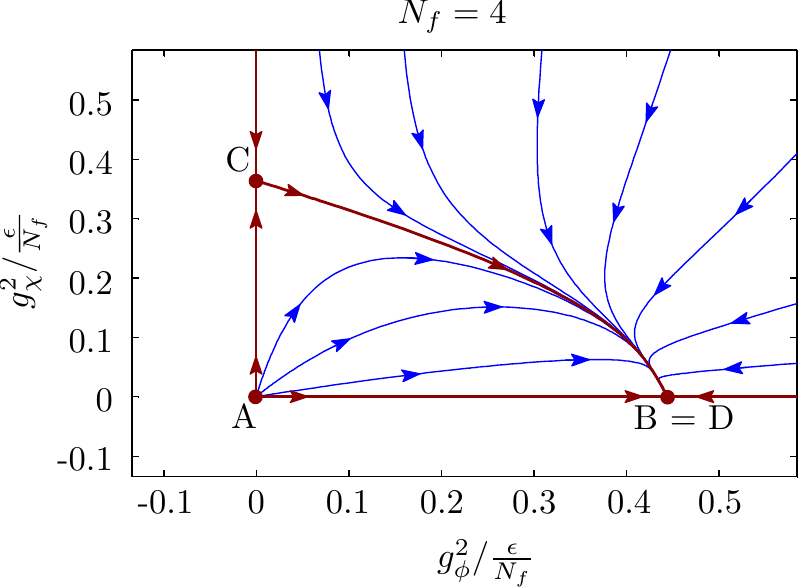} \\[\baselineskip]
 \includegraphics{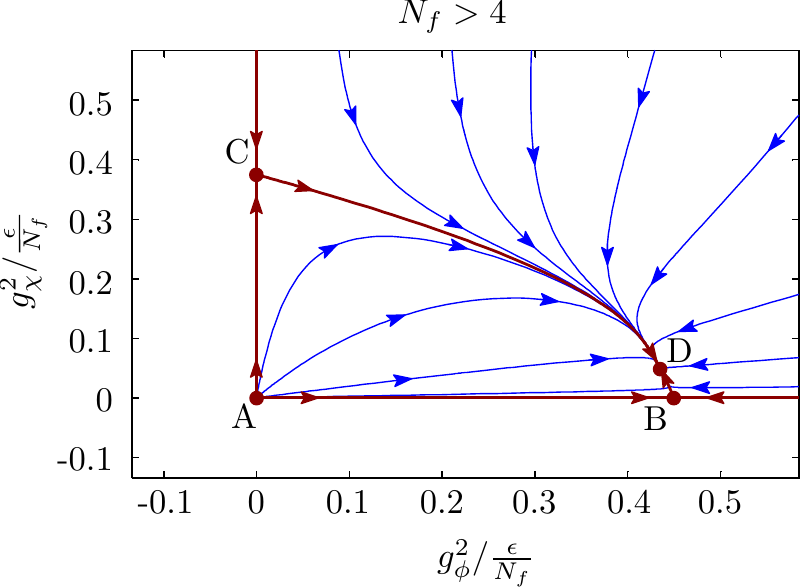} \hfill
 \includegraphics{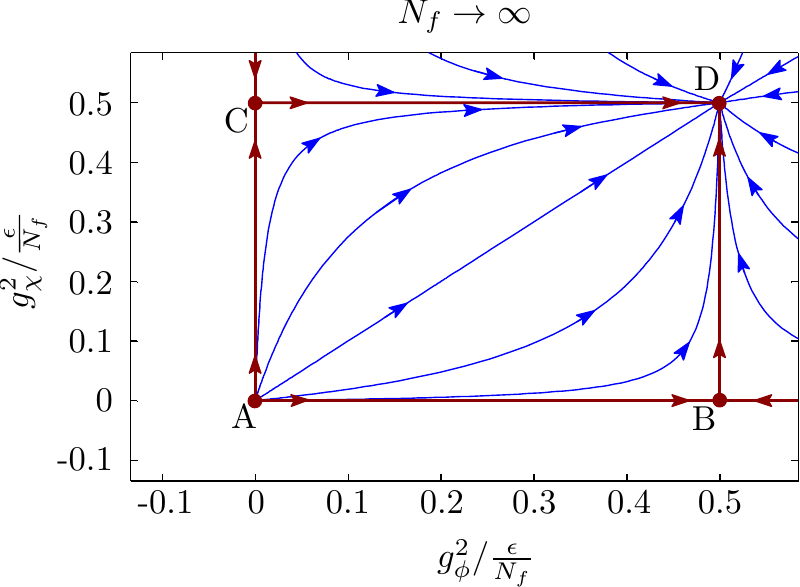}
 \caption{RG flow and fixed-point structure in the Yukawa-coupling subsector for different values of $N_f$ below, at, and above the critical value of $N_f = 4$. For $N_f<4$ fixed point B is stable within this subsector, while fixed point D is located in the unphysical domain  $g_\chi^2 < 0$ (top left). At $N_f=4$, D becomes physical and collides with B (top right), and subsequently exchanges stability with the latter for $N_f > 4$ (bottom left). For $N_f \rightarrow \infty$ the stable fixed point D is located at $g_\phi^{*2} = g_{\chi}^{*2}$ (bottom right). Note that A, B, C, and D may be true (and possibly stable) fixed points of the full system only if suitable corresponding zeros of the bosonic beta functions can be found, see text.}
\label{fig:flow}
\end{figure*}

The flow in the $g_\phi^2$-$g_\chi^2$ plane decouples from the bosonic flow equations, and analytical solutions of the zeros of the Yukawa-coupling beta functions $\{\beta_{g_\chi^2}, \beta_{g_\phi^2}\}$ can be readily displayed.
We find the values
\begin{align}
	\text{A:}&\ g_\chi^{2,\ast}=0, \ g_\phi^{2,\ast}=0, \\
	\text{B:}&\ g_\chi^{2,\ast}=0, \ g_\phi^{2,\ast}=\frac{\epsilon}{1+2 N_f}, \\
	\text{C:}&\ g_\chi^{2,\ast}=\frac{\epsilon}{3+2 N_f}, \ g_\phi^{2,\ast}=0, \\
	\text{D:}&\ g_\chi^{2,\ast}=\frac{\frac{1}{2}(N_f-4)\epsilon}{N_f^2+2 N_f-6},\ g_\phi^{2,\ast}=\frac{\frac{1}{2} N_f\epsilon}{N_f^2+2 N_f-6}. \label{eq:g-D}
\end{align}
We illustrate the flow and fixed-point structure in the $g_\phi^2$-$g_\chi^2$ plane for different values of $N_f$ in Fig.~\ref{fig:flow}. Let us emphasize, however, that the zeros A, B, C, and D, in order to represent true fixed points of the full system, need to be supplemented with suitable (and physically admissible) fixed-point values in the bosonic sector. In terms of the rescaled bosonic couplings $\{\lambda_\chi, \lambda_\phi, \lambda_{\chi\phi}\}\rightarrow \{\lambda_\chi/(8\pi^2\Lambda^\epsilon), \lambda_\phi/(8\pi^2\Lambda^\epsilon), \lambda_{\chi\phi}/(8\pi^2\Lambda^\epsilon)\}$ the $\beta$ functions in the bosonic sector read
\begin{align}
	\frac{d\lambda_\chi}{d\ln b}&=\epsilon \lambda_\chi-36\lambda_\chi^2-3\lambda_{\phi\chi}^2-4 N_f \lambda_\chi g_\chi^2+N_f g_\chi^4\,,\\
	\frac{d\lambda_\phi}{d\ln b}&=\epsilon \lambda_\phi -44\lambda_\phi^2-\lambda_{\phi\chi}^2-4 N_f \lambda_\phi g_\phi^2+N_f g_\phi^4\,,\\
	\frac{d\lambda_{\phi\chi}}{d\ln b}&=\epsilon \lambda_{\phi\chi}-8\lambda_{\phi\chi}^2 -20\lambda_\phi\lambda_{\phi\chi}-12\lambda_\chi\lambda_{\phi\chi}\nonumber\\
	&\quad-2 N_f \lambda_{\phi\chi} g_\phi^2-2 N_f \lambda_{\phi\chi} g_\chi^2+6 N_f g_\chi^2 g_\phi^2\,.
\end{align}

For a fixed point A, B, C or D to be physically admissible it has to satisfy a number of conditions.
First, the square of the Yukawa couplings has to be positive, $g_i^2 \geq 0, i \in \{\phi, \chi\}$. Also we need $\lambda_\chi, \lambda_\phi\geq 0$ to have an action that is bounded from below.
This is accompanied by the condition that $\lambda_{\phi\chi}$ is not too negative, i.e. if $\lambda_{\phi\chi}<0$ we need $4\lambda_\chi\lambda_\phi-\lambda_{\phi\chi}^2>0$.
Further, we will have at least two relevant directions related to the mass parameters of the bosons $\chi$ and $\phi$. 
%
%
%
%
They correspond to the two tuning parameters in the phase diagram in Fig.~\ref{fig:phasediag}.
This implies that the physical fixed point may not have more than these two relevant directions,
in order to be accessible without tuning a third microscopic parameter. Such fixed points will be called here ``stable''.
We hence classify the fixed points according to their number of relevant directions, i.e., their number of positive eigenvalues of the stability matrix.
For the fully Gaussian fixed point A, we find that the Yukawa couplings already provide two additional relevant directions with eigenvalues $\epsilon$ for any $N_f$. Thus in the following it will be discarded in our discussion.

Fixed point B leads to the least number of relevant directions in the full system when supplemented with the bosonic couplings
\begin{align}
\lambda_\chi^\ast&=\frac{\epsilon}{36},\quad \lambda_{\phi\chi}^\ast=0\,,\\
\lambda_\phi^\ast&=\frac{1-2 N_f+\sqrt{1+4 N_f(43+N_f)}}{88(2 N_f+1)}\epsilon,
\end{align}
which, in fact, means that the order parameters $\phi$ and $\chi$ decouple at B, with the Heisenberg field at its fermionic \emph{chiral} fixed point, and the Ising field at the purely bosonic (standard) Ising fixed point. We thus denote fixed point B also as ``chiral Heisenberg plus Ising'' (cH+I) fixed point. Its critical exponents are
\begin{align}
\theta_1&=2-\frac{\epsilon}{3}\,,\\
\theta_2&=2-\frac{5+34 N_f+5\sqrt{1+4N_f(43+N_f)}}{22(2 N_f+1)}\epsilon\,,\\
\theta_3&=\frac{N_f-4}{N_f+\frac{1}{2}}\epsilon\label{eq:theta3cHI}\,.
\end{align}
For the physical situation with $N_f=2$ we obtain $\{\theta_1,\theta_2,\theta_3\}=\{2-\frac{\epsilon}{3},2-\frac{84\epsilon}{55},-\frac{4\epsilon}{5}\}$, rendering the cH+I fixed point stable as $\theta_3<0$.
The situation changes upon increasing the value of $N_f$, cf.\ Fig.~\ref{fig:flow}. For $N_f = 4$, B collides with another fixed point D, with which it exchanges stability for $N_f > 4$.  For $N_f > 4$ the cH+I fixed point thus becomes unstable. The scenario is similar to the well-known situation in the purely bosonic $\mathrm O(N)$ models, in which the Wilson-Fisher fixed point approaches the Gaussian fixed point when the dimension is increased towards the upper critical dimension, and subsequently exchanges stability with the latter.

Fixed point C, when completed with the bosonic fixed-point values
\begin{align}
\lambda_\chi^\ast&=\frac{3-2 N_f+\sqrt{9+4 N_f(33+N_f)}}{72(3+2 N_f)}\epsilon\,,\\
\lambda_\phi^\ast&=\frac{\epsilon}{44},\quad \lambda_{\phi\chi}^\ast=0,
\end{align}
could then, within our nomenclature, be termed ``chiral Ising plus Heisenberg'' (cI+H) fixed point. It has {\it three} relevant directions for all admissible choices of $N_f$ and $\epsilon$,
\begin{align}
\theta_1&=2-\frac{5\epsilon}{11}\,,\\
\theta_2&=2-\frac{3+10 N_f+\sqrt{9+4 N_f(33+N_f)}}{6(3+2 N_f)}\epsilon\,,\\
\theta_3&=\frac{2 N_f}{3+2 N_f}\epsilon\,,
\end{align}
and is thus never stable. Its critical exponents and anomalous dimensions for the physical case $N_f=2$ are listed together with the those of fixed point B in Table \ref{Tab:eps_Ising}.
\begin{table}[tb]
\caption{\label{Tab:eps_Ising} The largest three critical exponents and anomalous dimensions for fixed points B and C in first order $\epsilon$ expansion for the physical choice $N_f=2$. Here, fixed point B (``chiral Heisenberg plus Ising'' fixed point) is stable and thus governs the multicritical behavior. In the decoupled bosonic sector the anomalous dimension vanishes as it is of $\mathcal{O}(\epsilon^2)$.}
\begin{tabular*}{.48\textwidth}{@{\extracolsep{\fill} } c | c c c | c c c}\hline\hline
& $\theta_1$ & $\theta_2$&$\theta_3$ & $\eta_\phi$ & $\eta_\chi$ & $\eta_\psi$\\
\hline
B (cH+I) & $2-\frac{1}{3}\epsilon$ & $2-\frac{84}{55}\epsilon$ & $-\frac{4}{5}\epsilon$  & $\frac{4}{5}\epsilon$  & 0 &$\frac{3}{10}\epsilon$ \\
C (cI+H) & $2-\frac{5}{11}\epsilon$  & $2-\frac{20}{21}\epsilon$ & $\frac{4}{7}\epsilon$ & 0 & $\frac{4}{7}\epsilon$ & $\frac{1}{14}\epsilon$ \\
\hline\hline
\end{tabular*}
\end{table}

\begin{figure}[t!]
\centering
 \includegraphics[width=.48\textwidth]{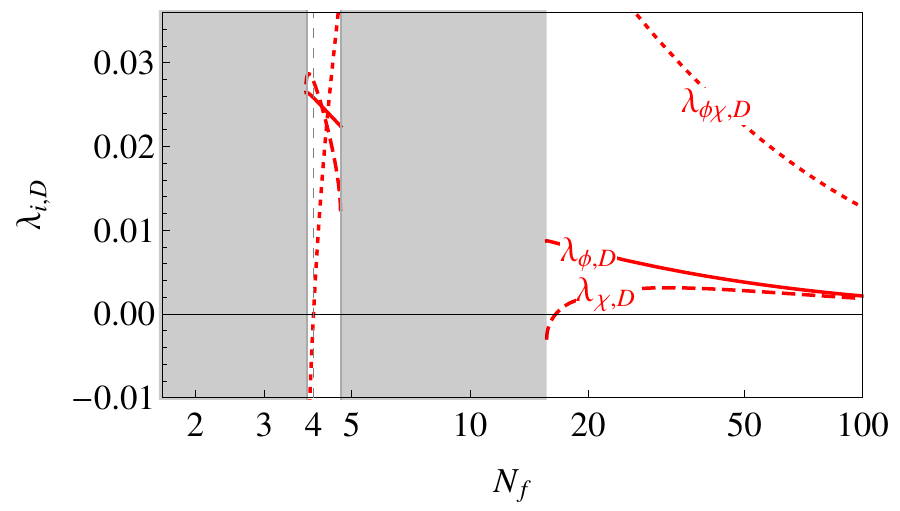}
 \includegraphics[width=.48\textwidth]{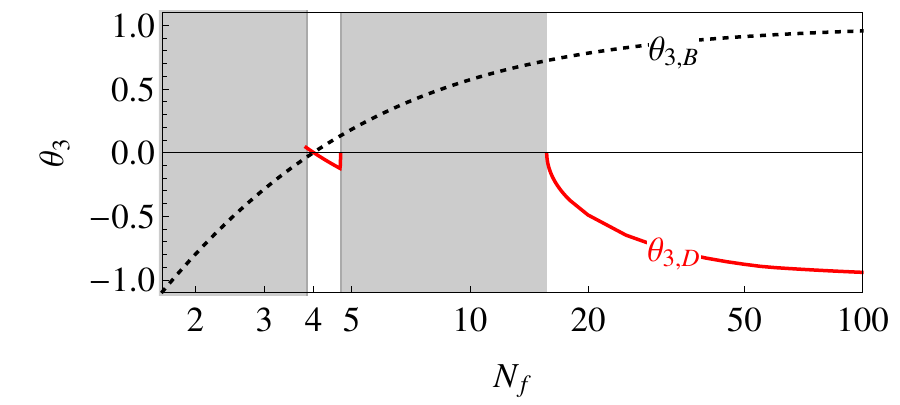}
 \caption{Upper panel: Coordinates of the bosonic couplings at the fixed point D as a function of $N_f$. The gray areas indicate the ranges of $N_f$ where fixed point D does not exist and no other stable fixed point exists.  Lower panel: Third critical exponent as function of $N_f$ for the two possible stable fixed points B and D.}
\label{fig:FPD}
\end{figure}

For the mixed fixed point D the computation of the bosonic fixed-point values $\lambda_i^*$ is slightly more involved, but can readily be done numerically. The bosonic fixed-point couplings lead to a quite intriguing behavior as a function of $N_f$, as displayed in Fig.~\ref{fig:FPD}.
First, we observe that there are two ranges of $N_f$ where no real fixed-point values in the bosonic sector can be found for the given values of $g_\phi^{*2}$ and $g_\chi^{*2}$ [Eq.~\eqref{eq:g-D}].
The first range covers small $N_f\lesssim 3.8$ and the second range is $N_f \in [4.7,15.7]$.
These intervals are indicated in Fig.~\ref{fig:FPD} by the gray-shaded areas.
In the narrow range $N_f \in [3.8,4.7]$ the solution for fixed point D is physically admissible only for $N_f\geq4$, since $g_\chi^2<0$ for $N_f<4$. For $N_f=4$ the fixed points B and D collide and exchange stability, so that for $4.7 >  N_f\geq4$ the fixed point D is stable.
For $15.7 \lesssim N_f \lesssim 16.6$, while corresponding real fixed-point values can now be found, the solution for fixed point D remains unphysical, since $\lambda_\chi<0$ (see Fig.~\ref{fig:FPD}).
Finally, for $N_f \gtrsim 16.6$ the fixed point D constitutes a physically admissible solution with negative exponent $\theta_3$, i.e., it is the stable fixed point.
We have checked that between $N_f=4$ and $N_f=16$ no other real and stable solution exists in our approximation scheme. In the limit of large $N_f$, we find
\begin{align}\label{eq:largeNf}
g_\chi^{2,\ast}= g_\phi^{2,\ast}=\frac{\epsilon}{2N_f}\,,
\end{align}
and we obtain a stable fixed point in agreement with the large-$N_f$ calculation~\cite{herbut2006}. It has the bosonic coupling coordinates
\begin{align}
\lambda_\chi^\ast=\lambda_\phi^\ast=\frac{\epsilon}{4 N_f}, \quad \lambda_{\phi\chi}^\ast=\frac{3\epsilon}{2 N_f}\,.
\end{align}

%


\section{Discussion}
\label{sec:discussion}

\begin{figure}[t!]
\centering
 \includegraphics[width=0.48\textwidth]{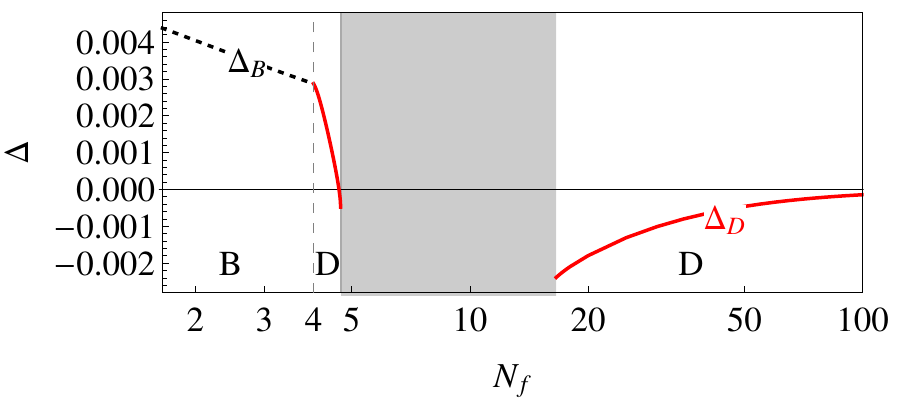}
 \caption{Value of the parameter $\Delta$ as a function of $N_f$ for the stable fixed point, which governs the multicritical behavior in the phase diagram of Fig.~\ref{fig:phasediag}(a). For $N_f \leq 4$ the multicritical point is given by the chiral Heisenberg plus Ising fixed point (B) and it has positive $\Delta > 0$, corresponding to tetracritical behavior with a mixed CDW and SDW phase, as indicated in Fig.~\ref{fig:phasediag}(b). For $4<N_f<4.64$ the multicritical point is still tetracritical, but its long-range behavior defines a new universality class governed by fixed point D. For $4.64<N_f<4.7$ the multicritical point is bicritical (see Fig.~\ref{fig:phasediag}(c)). For $4.7<N_f<16.6$ there is no physically admissible stable fixed point in the system and we expect the multicritical point to become discontinuous. For $N_f>16.6$ the multicritical behavior is again bicritical.}
\label{fig:Delta}
\end{figure}

An important quantity to classify the critical behavior of a multicritical point is the parameter $\Delta$,
\begin{align}
\Delta=4\lambda_\chi\lambda_\phi-\lambda_{\phi\chi}^2,
\end{align}
whose sign determines the nature of the multicritical point, i.e., whether it is bicritical or tetracritical.
 When $\Delta \leq 0$ the transition line between the two ordered phases is expected to be of first order, whereas for  $\Delta >0$ coexistence of the two orders is expected with second order transitions between the four different regimes.
This conclusion is based on a consideration of the saddle-point approximation of the free energy~\cite{calabrese2003,kivelson2001}, and we expect
that the presence of Dirac fermions does not affect it, since the argument relies only on the consideration of the boson effective potential.
We identify three different regimes if we classify the nature of the multicritical point in terms of $\Delta$, cf. Fig.~\ref{fig:Delta}.
For small fermion flavor number $N_f\lesssim4.64$, including the graphene case $N_f=2$, we find a positive $\Delta$ at the stable fixed point.
Here, tetracritical behavior dominates the phase transitions and a mixed phase appears with a coexistence of SDW and CDW order.
In a large range of this regime $1\leq N_f\leq4$, the cH+I fixed point B with vanishing $g_\chi^2$ is stable indicating that the properties at the multicritical point are dominated by the chiral Heisenberg universality class.
This universality class also describes the transition from the semimetal to the antiferromagnetic state for increasing nearest-neighbor interaction.
The stability suggests the continuation of this behavior up to the multicritical point.
On the bosonic side of the transition the order parameters decouple, whereas they interact with each other if fixed point D becomes stable.
The stability of fixed point D is accompanied by a negative $\Delta$ for $N_f\in(4.64,4.7)$ and $N_f>16.6$.
In this regime the phase diagram exhibits bicritical behavior with continuous transitions between SM and SDW or CDW, respectively, and a first order line between SDW and CDW. The large-$N_f$ behavior is thus consistent with the previous $1/N_f$ approaches~\cite{herbut2006}.
The intermediate regime $N_f\in(4.7,16.6)$ is characterized by the absence of any stable fixed point.
Here, the second order lines of the separate transitions from SM to SDW and CDW end in a discontinuous point and the transition from SDW to CDW is also of first order.

In conclusion, we presented an effective description of the multicritical point between the semimetallic, the CDW and the SDW state on the honeycomb lattice that becomes exact near three spatial dimensions.
 A rather complex picture emerges as a function of the fermion flavor number, in which the graphene case appears dominated by the chiral Heisenberg universality class, at least to the present order of the $\epsilon$ expansion.
A tetracritical phase diagram with a coexistence phase of SDW and CDW is the consequence.
In purely bosonic multicritical systems \cite{nelson1974,calabrese2003}, as well as in the separate Gross-Neveu-Yukawa models \cite{janssen2014}, 
the leading-order $\epsilon$ expansion
captures well the qualitative behavior of the systems.
We expect this to hold for our model as well
and, at the very least, our results should qualitatively correct describe the $N_f$ dependence of the fixed-point structure in two spatial dimensions as appropriate for the description of graphene.
However, the bare numbers of the critical $N_f$ values
at which the multicritical behavior of the system changes can possibly be reduced substantially if one went beyond the leading-order $\epsilon$ expansion \cite{herbut1997,fei} and more elaborate investigations may be needed to settle the true nature of the multicritical point in graphene's phase diagram.
A promising complimentary approach to the critical $N_f$ values could be provided by a functional renormalization group study \cite{eichhorn2013,janssen2014} which allows to work directly in two spatial dimensions.
This is under way~\cite{classen}.

\acknowledgments
We acknowledge discussions with I. Boettcher, A. Eichhorn and L. von Smekal. L.C. acknowledges support by the Studienstiftung. M.M.S. thanks the Theory Division at CERN for hospitality during the course of this work. M.M.S. is supported by the grant ERC-AdG-290623 and DFG grant FOR 723. I.F.H and L. J. were supported by the NSERC of Canada. L. J. was also supported by the DFG,  JA\,2306/1-1.


\end{document}